\documentclass[reprint,aps,prd,nofootinbib]{revtex4-1}
\usepackage{hyperref}
\usepackage[utf8]{inputenc}
\usepackage[utf8]{inputenc}
\usepackage{amsmath}
\usepackage{amsfonts}
\usepackage{amssymb}
\usepackage[left=2cm,right=2cm,top=2cm,bottom=2cm]{geometry}
\usepackage{braket}
\usepackage{dsfont}

\usepackage{graphics}
\usepackage{tikz}
\usetikzlibrary{arrows,decorations.pathmorphing,backgrounds,positioning,fit,petri,shapes,snakes}
\usepackage{lipsum}

\usepackage{multirow}
\usepackage{verbatim}
\begin{document}

\title{Testing the Cosmic Shear Spatially-Flat Universe Approximation with Generalized Lensing and Shear Spectra}

\begin{abstract}
We introduce the Generalised Lensing and Shear Spectra ({\tt GLaSS}) code which is available for download from \url{https://github.com/astro-informatics/GLaSS}
It is a fast and flexible public code, written in {\tt Python}, that computes generalized spherical cosmic shear spectra.
The commonly used tomographic and spherical Bessel lensing spectra come as built-in run-mode options. {\tt GLaSS} is integrated into the {\tt Cosmosis} modular cosmological pipeline package.
We outline several computational choices that accelerate the computation of cosmic shear power spectra. 
Using {\tt GLaSS}, we test whether the assumption that using the lensing and projection kernels for a spatially-flat universe -- in a universe with a small amount of spatial curvature -- negligibly impacts the lensing spectrum.
We refer to this assumption as The Spatially-Flat Universe Approximation, that has been implicitly assumed in all cosmic shear studies to date.
We confirm that The Spatially-Flat Universe Approximation has a negligible impact on Stage IV cosmic shear experiments.
\end{abstract}

\author{Peter L. Taylor}
\email{peterllewelyntaylor@gmail.com}
\author{Thomas D.~Kitching}
\author{Jason D.~McEwen}
\affiliation{Mullard Space Science Laboratory, University College London, Holmbury St.~Mary, Dorking, Surrey RH5 6NT, UK}
\author{Thomas Tram}
\affiliation{Department of Physics and Astronomy, University of Aarhus, Ny Munkegade 120, DK–8000 Aarhus C, Denmark}
\affiliation{Aarhus Institute of Advanced Studies (AIAS), Aarhus University, DK–8000 Aarhus C,
Denmark}
\date{12 March 2018}
\maketitle

\section{Introduction}
The shape of distant galaxies is distorted by inhomogeneities in the gravitational field along the line of sight; a phenomenon known as gravitational lensing. When the distortion is small, as is most commonly the case, the change in shape is a change in the size and ellipticity of the observed image; known as shear. The gravitational lensing caused by large-scale structure, and in particular the two-point correlation function or power spectrum of this effect, is called cosmic shear.
\par Experiments that measure cosmic shear are sensitive to the physics of the late Universe, making them an ideal probe to distinguish between models of dark energy \cite{refregier2010euclid}. Stage IV weak lensing experiments, that include Euclid\footnote{\url{http://euclid-ec.org}} \cite{laureijs2010euclid}, WFIRST\footnote{\url{https://www.nasa.gov/wfirst}} \cite{spergel2015wide} and LSST\footnote{\url{https://www.lsst.org}}~\cite{anthony4836large}, will provide an order of magnitude improvement in the precision and accuracy of cosmological parameter estimation over existing surveys \cite{albrecht2006report}.
\par To prepare for these upcoming experiments we must prepare fast and accurate codes to compute the theoretical cosmic shear power spectra for any cosmology. While there are already publicly available tomographic lensing codes that use the Limber approximation \cite{kilbinger2009dark,cosmosis}, there are no other codes that can compute the cosmic shear power spectra with an arbitrary weight function. It remains an open question which weight-function optimally extracts cosmological information, and we leave this for a future work.  
\par Also, before the arrival of Stage IV data, it is vital to test the validity of all assumptions used in cosmic shear studies. One of these approximations is that for the purposes of computing the cosmic shear power spectra we can always treat the Universe as spatially flat. This is an assumption that has not been tested previously.   
\par The structure of this paper is as follows. In Section \ref{sec:formalism} we review the equations for the cosmic shear power spectra and the effect of spatial-curvature on the lensing kernel and projection kernel. In Section \ref{sec:code} we introduce {\tt GLaSS}, which computes lensing spectra, and discuss a few computational choices that we implemented to speed up the computation of cosmic shear power spectra. Finally in Section \ref{sec:results} we demonstrate the speed of {\tt GLaSS} and discuss the impact of the Spatially-Flat Universe Approximation.

\section{Formalism} \label{sec:formalism}
\subsection{Generalized-Spherical Lensing Spectra}
The \textit{generalized spherical-transform} is defined in \cite{taylor2018preparing}: 
\begin{equation} \label{eq:C}
\gamma_{\ell m} \left( \eta \right) =  \sqrt{\frac{2}{\pi}} \sum _g \gamma_g \left( r_g, \boldsymbol{\theta_g } \right)W_{\ell} \left(\eta, r_g \right)  {}_2 Y_{\ell m} \left( \boldsymbol {\theta_g} \right), 
\end{equation}

where $\gamma \in \mathbb{C}$ is the shear, the sum is over all galaxies $g$ with angular coordinate ${\bf \theta_g}$ and radial coordinate $r_g$, $W_\ell$ is a weight and $_2 Y_{\ell m}$ are the spin-$2$ spherical harmonics. The cosmic shear power spectrum in this basis is:
\begin{equation} \label{eq:c_l}
C_{\ell}^{\gamma \gamma} \left( \eta_1, \eta_2 \right) =  \frac{9 \Omega_m ^ 2 H_0 ^ 4}{16 \pi^4 c^ 4 }\frac{\left( \ell + 2 \right)!}{\left( \ell - 2 \right)!} \int \frac{\text{d} k}{ {k} ^2} G_{\ell}^\gamma \left( \eta_1, k \right) G_{\ell}^\gamma \left(\eta_2, k \right) ,
\end{equation}
where $\Omega_m$ is the fractional energy density of matter, $c$ is the speed of light in vacuum and $H_0$ is the value of the Hubble constant today. The $G$-matrix is:
\begin{equation} \label{eq:G}
\begin{aligned}	
G_{\ell} ^ \gamma \left( \eta , k \right) \equiv \int \text{d}z_p \text{d} z' \text{ }  &n \left(z_p \right) p \left(z' | z_p \right) \\ & \times W_{\ell} \left(\eta, r  \left[z' \right]\right) U_{\ell} \left(r \left[ z' \right], k \right)  
\end{aligned}
\end{equation}
where $r[z]$ is the co-moving distance at a redshift $z$ and the $U$-matrix is:
\begin{equation} \label{eq:U}
U_{\ell} \left(r[z], k \right) \equiv \int ^ r _0 \text{d} r' \text{ } \frac{F_K \left(r, r' \right)}{a \left(r' \right)} j_{\ell} \left( k r' \right) P^{1/2 }\left(k ; r' \right),
\end{equation}
where $a$ is the scale factor, $j_\ell(kr)$ are the spherical Bessel functions and $P(k;r)$ is the power spectrum. The radial distribution of galaxies is denoted by $n(z)$ and $p \left( z|z' \right)$ gives the probability that a galaxy has a redshift $z$, given a photometric redshift measurement $z'$. For a spatially-flat cosmology the lensing kernel, $F_K \left(r, r' \right)$, is:
\begin{equation} \label{eq:kernel}
 F_K \left(r, r' \right)\equiv \frac{r-r'}{rr'}.
\end{equation} 
The power spectrum caused by the random ellipticity component of galaxies, the shot noise spectrum, is given by:
\begin{equation} \label{eq:Noise}
N_\ell^{ e e} \left( \eta_1, \eta_2 \right) = \frac{\sigma_e ^2}{2 \pi ^ 2} \int \text{d} z \text{ } n\left( z \right)W_\ell \left(\eta_1, r \right)W_\ell \left(\eta_2, r \right) , 
\end{equation}
where $\sigma_e ^2$ is the variance of the intrinsic (unlensed) ellipticities of the observed galaxies. We take $\sigma_e  = 0.3$ throughout \cite{brown2003shear}.
\par Taking the weight-function, $W_\ell \left(\eta, r  \left[ z \right]\right) \equiv j_\ell \left(\eta r[z]\right)$ in equations~\eqref{eq:G} and~\eqref{eq:Noise} yields the equations for `3D cosmic shear' first proposed in \cite{heavens3d}. To recover the `tomographic' cosmic shear spectra, first proposed in \cite{hushotnoise}, we take the weight function, $W^I$, as a top hat function in redshift only:
\begin{equation}
   W ^ I \left(z \right) \equiv
    \begin{cases}
      1 & \text{if $z \in I$  }\\
      0 & \text{if $z \notin I$,  }\\
    \end{cases} 
\end{equation}
the tomographic bin associated with redshift region $I$.
\par Taking the Limber approximation ~\cite{loverdelimber}, the $U$-matrix becomes:
\begin{equation} \label{eq:limber}
U_{\ell} \left(r, k \right) = \frac{F_k \left( r, \nu \left( k \right) \right)}{k a \left( \nu\left( k \right)  \right)} \sqrt {\frac{\pi}{2 \left( {\ell} + 1/2 \right)}}  P ^ {1/2} \left( k, \nu\left( k \right)  \right),
\end{equation}
where $\nu\left( k \right)  \equiv \frac{{\ell}+ 1/2}{k}$. This is a good approximation for $\ell > 100$~\cite{kitchinglimits,kilbinger2017precision}.

\subsection{The Lensing Kernel for $\Omega_k \neq 0$}
In a spatially-curved universe, the expression for the lensing kernel in equation~\eqref{eq:kernel} must be replaced by the more general expression:
\begin{equation}
 F_K \left(r, r' \right)\equiv \frac{f_k\left(r-r'\right)}{f_k\left(r\right) f_k\left(r' \right)}, 
\end{equation}
where $f_k(r)$ is the co-moving angular distance \cite{kilbingerreview}. This is given by:  
\begin{equation}
   f_K \left(r \right)  \equiv
    \begin{cases}
      K^{-1/2} \text{sin} \left(K ^ {1/2} r \right) & \text{if $K>0$  }\\
      r & \text{if $K=0$.  }\\
      \left(-K \right)^{-1/2} \text{sinh} \left(\left( -K \right) ^ {1/2} r \right) & \text{if $K<0$.  }\\
    \end{cases} 
\end{equation}
where the curvature, $K$, is defined as $K \equiv - \left( H_0 / c \right) ^ 2 \Omega_k$, and $\Omega_k$ is the spatial curvature density today. 

\subsection{The Projection Kernel for $\Omega_k \neq 0$}
In a spatially-flat universe, the gravitational potential at a time labeled by the redshift $z$, $\Phi \left( {\bf r}; z\right)$, is related to the underlying density field, $\delta \left( {\bf r};z\right)$, by the Poisson equation:
\begin{equation} \label{eq:poisson}
\nabla_r ^ 2 \Phi \left( {\bf r}; z \right) = \frac{3 \Omega_m H_0 ^ 2}{2 a\left( t\right)} \delta \left( {\bf r};z \right),
\end{equation}
where $\nabla ^2_r$ is the Laplacian associated with a spatially-flat universe. 
\par The potential, $\Phi \left( {\bf r};z\right)$, in the observer's frame is given in a coordinate system defined by two angles on the sky and a radial distance denoted by $(r, \theta, \phi)$. Meanwhile the density field is in rectilinear coordinates. To relate the two, and hence find the lensing spectra in terms of the matter power spectrum, we expand the potential in spherical Bessel space: 
\begin{equation} 
\Phi_{\ell m} \left( k \right) = \sqrt{\frac{2}{\pi}} \int \text{d}^ 3 r \text { } \Phi \left ( r \right) j_\ell \left( k r \right) Y_{\ell m} \left(\theta, \phi \right),
\end{equation}
where $j_\ell \left( k r \right)$ are spherical Bessel functions and $Y_{\ell m} \left(\theta, \phi \right)$ are spherical harmonics. Then since spherical harmonics and spherical Bessel functions are eigenfunctions of the Laplace operator, we have:
\begin{equation} \label{eq:eig}
\left( \nabla_r ^ 2 + k^2   \right) j_\ell \left(k r \right) Y_{\ell m} \left( \theta, \phi \right) = 0,
\end{equation}
and from equation~\eqref{eq:poisson} the lensing potential is related to the density field in harmonic space by:
\begin{equation}
 \Phi_{\ell m} \left( k; z \right) = - \frac{3 \Omega_m H_0 ^ 2}{2 k^2 a\left( t\right)} \delta_{\ell m} \left( k;z \right).
\end{equation}
From this it is possible to derive the expression for the cosmic shear power spectrum. Since Bessel functions relate the lensing potential in rectilinear coordinates to a projected shear signal on the sky, we refer to $j_\ell \left( kr \right)$ as the \emph{projection kernel}. In the final expression for the cosmic shear power spectra, the projection kernel is found in the $U$-matrix (see \cite{castro} for a full derivation).
\par Meanwhile in a spatially-curved universe, we must take the Laplacian associated with the curved Robertson-Walker metric~\cite{kosowsky1998efficient} in equation \eqref{eq:poisson}. Hence the projection kernel must change too. In particular spherical Bessel functions must be replaced by hyperspherical Bessel functions, $\Phi^\beta_l \left( r \right)$, because they are eigenfunctions of the the Laplace operator in a spatially-curved cosmology. That is:  
\begin{equation} \label{eq:curved_poisson}
\left( \nabla_{S_K(\chi)} ^ 2 + (ck)^2   \right) \Phi^\beta_\ell \left( \chi \right) Y_{\ell m} \left( \theta, \phi \right) = 0, 
\end{equation} \label{eq:beta_def}
where $\beta \equiv \sqrt{(ck)^2 + K}$, $\chi = r/c$ and
\begin{equation}
   S_K \left(\chi \right) \equiv
    \begin{cases}
      \sin{\chi} & \text{if $K > 0$  }\\
      \chi & \text{if $K = 0$  }\\
      \sinh{\chi} & \text{if $K < 0$.  }\\
    \end{cases} 
\end{equation}
Following the same argument used in the spatially-flat case, we find the hyperspherical Bessel functions enter the $U$-matrix, in place of the normal spherical Bessel functions, as the projection kernel.
\par The Limber approximation also has to be generalized to spatially-curved cosmologies \cite{lesgourgues2014fast}. In this case the Limber-approximated $U$-matrix becomes:
\begin{equation} \label{eq:modlimber}
U_\ell \left( r, k \right) = \left( 1 - \hat K \frac{\ell ^2 }{\beta ^ 2} \right) ^ {-\frac{1}{4}} U ^{\rm flat}_\ell \left( r, k \right),
\end{equation}
where $\hat K$ is the sign of the curvature $K$, and $U^{\rm flat}_\ell \left( r, k \right)$ is the Limber approximated $U$-matrix for a spatially-flat universe defined in equation \eqref{eq:limber}.

\section{The GLaSS Code} \label{sec:code}
We now describe the {\tt GLaSS} code that can compute all the power spectra previously described. 
\subsection{Description and Run Options}
{\tt GLaSS} is a flexible code written in {\tt Python} and it is fully integrated into the {\tt Cosmosis} modular cosmological pipeline~\cite{cosmosis}. The code is provided with {\tt Python} wrappers and cosmological information can be read directly from the {\tt Cosmosis} pipeline or from an external source. 
\par There are numerous run-mode options. The user can choose between several weights. These include: the top hats associated with tomographic binning with an equal number of galaxies per bin or equally spaced tomographic bins in redshift, the spherical Bessel weight, or a customized weight provided by the user. The number of tomographic bins can also be varied.  The user can specify which $\ell$-modes to sample over a prescribed redshift range. The package is distributed with default functional forms for the radial distribution of galaxies, $n(z)$, and photometric redshift error $p \left( z|z' \right)$. These are:
\begin{equation} \label{eq:photo error}
p \left( z | z_p \right) \equiv \frac{1}{2 \pi \sigma_z \left(z_p \right)} e ^{- \frac{ \left( z -c_{\rm cal} z_p + z_{\rm bias} \right) ^2  } {2 \sigma_{z_p}} },
\end{equation}
with $c_{\rm cal} = 1$, $z_{\rm bias} = 0$ and $\sigma_{z_p} = A \left(  1 + z_p \right)$, with default value is $ A = 0.05$ \cite{ilbert2006accurate} and
\begin{equation} \label{eq:n(z)}
n \left( z_p \right) \propto \frac{a_1}{c_1} e ^ {- \frac{ \left( z-0.7 \right) ^2 }{b_1^2} } + e ^ {- \frac{ \left( z-1.2 \right) ^2 }{d_1^2} } ,
\end{equation}
with default values $\left(a_1/c_1,b_1 ,d_1 \right) =\left( 1.5 / 0.2, 0.32, 0.46 \right)$ \cite{van2013cfhtlens}. It is possible for the user to provide custom functional forms too. 
\par The Limber approximation can be turned on or off. Since the Limber approximation is less accurate at low-$\ell$ \cite{kitchinglimits}, it can be turned on for any chosen $\ell > \ell_{\rm Lim}$, for a specified value of $\ell_{\rm Lim}$.
\par Finally it is possible to independently turn the spatially-curved lensing kernel and projection kernel approximations on or off; however later we show these approximations have negligible impact. Hyperspherical Bessel functions are computed with a {\tt Python} wrapper that calls {\tt CLASS} \cite{blas2011cosmic}. Details about the implementation of the hyperspherical Bessel functions in {\tt CLASS} are given in (\cite{tram2017computation} and \cite{lesgourgues2014fast}). 
\par {\tt GLaSS} has been compared to the spherical Bessel code used in~\cite{spuriomancini} and gives very similar output when using the spherical Bessel weight (Spurio Mancini et al. in prep).

\subsection{Computational Choices}
Several numerical choices have been implemented in {\tt GLaSS} to reduce the computation time. 
\par Values of the Bessel functions, $j_\ell \left(x \right)$, are computed just once and stored in a 2D look up table in $\ell$ and $x$.  The values of $j_\ell \left(k r \right)$, can then be found as needed. We sample sufficiently densely in $x$ so that final lensing spectra is not affected above machine precision. Compressing the data in this way reduces memory requirements and was used before in~\cite{seljak469line, kosowsky1998efficient}. In the hyperspherical case, it is not possible to compress the data to a 2D-array. In this case the hyperspherical Bessel functions are computed on the fly, slowing down the total computation time.
\par Even though the Bessel functions need only be computed once, the computation of these has also been optimized in {\tt GLaSS}. For a given argument $x$, {\tt GLaSS} computes and stores all $j_\ell (x)$ for all $\ell$-modes simultaneously using Miller's algorithm which is based on recurrence relations and implemented in the {\tt GNU Scientific Library} \cite{gnu}, and called using {\tt ctypesGSL}. If the maximum $\ell$ is too high, Miller's algorithm suffers from underflow. {\tt GLaSS} avoids this by first sparsely sampling the $x$-range to determine a maximum $\ell_\text{max} \left( x\right)$, for each $x$, which is defined as the $\ell$-value past which the Bessel functions fall below machine precision. {\tt GLaSS} sets $j_\ell (x) = 0$ for all $\ell > \ell_\text{max} \left( x\right)$.
\par As the Bessel functions are pre-computed, the majority of the computation time is taken by evaluating the nested integrals in equations~\eqref{eq:C} - \eqref{eq:U}. In {\tt GLaSS} all these are evaluated using  matrix multiplications on a grid in $r$ and $k$. For example the $U$-matrix can be written as a matrix multiplication given by:
\begin{equation} \label{eq:mat}
U_{\ell} \left(r, k \right) \approx   \sum_{r'} A \left(r, r' \right)B \left(r', k \right),
\end{equation}
$A \left( r, r' \right) \equiv \Delta r' \frac{F_K \left(r, r' \right)}{a \left(r' \right)} $, where $\Delta r'$ is the spacing of the grid in $r'$ and $B \left( r, r' \right) \equiv  j_{\ell} \left( k r' \right) P^{1/2 }\left(k ; r' \right).$ 
\par All matrix multiplications in {\tt GLaSS} are implemented using the {\tt numpy.dot} function. This is one of the few functions that releases the Global Interpreter Lock in {\tt Python}, so the matrix multiplications are parallelized when {\tt numpy} is linked to a linear algebra library such as {\tt BLAS (Basic Linear Algebra Subprograms)}, {\tt Math Kernel Library (MKL)} or {\tt Apple Accelerate}. There are also {\tt MPI} run-mode options for the Monte Carlo samplers in {\tt Cosmosis}, which can be used to further distribute the workload over multiple cores.    
\par The final speed improvements come from making the Limber approximation. Since the Bessel functions oscillate quickly, particularly for high-$\ell$, making the Limber approximation reduces the size of the computation grid needed to accurately evaluate the $U$-matrix. Meanwhile {\tt GLaSS} can simultaneously turn the Limber approximation off at low-$\ell$ so that accuracy is not lost at these large angular scales where the Limber approximation is invalid. 

\section{Results} \label{sec:results}
We now present results on the {\tt GLaSS} computational scaling, and the impact of the spatially-flat universe approximation. In what follows we assume a 15,000 square degrees survey with 30 galaxies per $\text{arcmin} ^2$ as predicted for the Euclid wide-field survey. 
\subsection{GLaSS Module Timing}
   \begin{figure}
   \centering
    \vspace{2mm}
    \includegraphics[width=85mm]{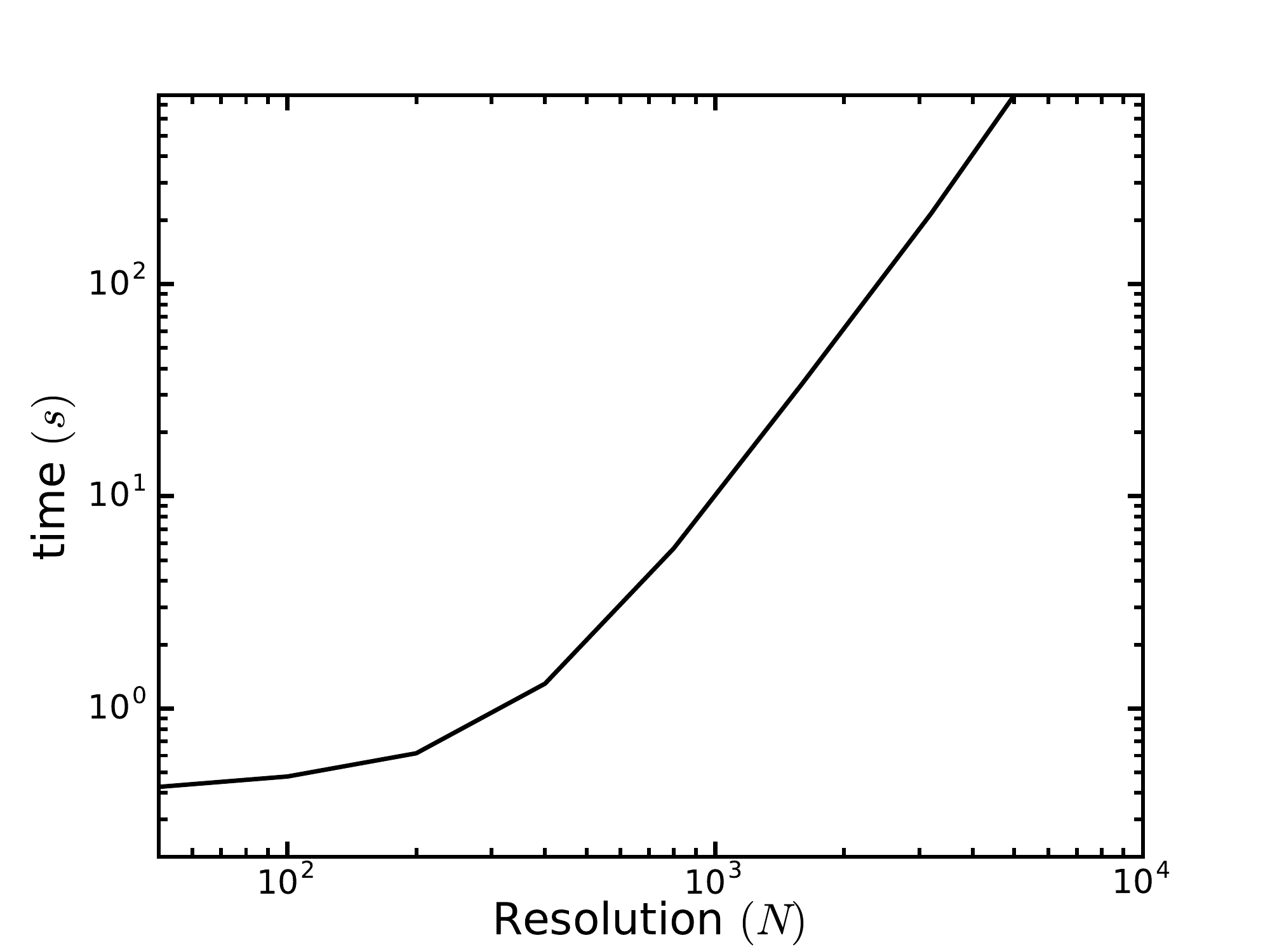}
    \caption{Time to compute $50$ $\ell$-modes in {\tt GLaSS} against grid  resolution $N$ (i.e. number of tomographic bins, or $k$-modes in 3D cosmic shear) on a single { \tt 2.7 GHz Intel i5 Core}. Spectra for resolutions up to $N = 400$ can all be computed in less than $1.5s$. This is a sufficient resolution to evaluate the tomographic cosmic shear power spectra and recover all information from the 3D shear field \cite{taylor2018preparing}. At high-$\ell$, the computation time, $t$, is dominated by the matrix multiplications and hence  scales as $t \propto N ^{2.87}.$}
    \label{fig:timing}
    \end{figure}

We now present the results of several speed tests using {\tt GLaSS}. All results cited are for $10$-bin tomography with an equal number of galaxies per bin sampling $50$ $\ell$-modes below $\ell_\text{max} = 3000$ on a single {\tt 2.7 GHz Intel i5 Core} on a 2015 Macbook Pro with $8$ GB of RAM. 
\par It takes 28 seconds to compute all the Bessel function data, but this must only ever be computed once. This shows how vital it is to pre-compute the Bessel data.
\par The nested integral and hence the lensing spectra are computed on an $N \times N$ grid, where $N$ is the number of linearly sampled points in $z$ and logarithmically sampled points in $k$. For $N < 300$, it takes less than a second to compute the lensing spectra when the Limber approximation is assumed for $\ell > 100$.
\par As the resolution is increased beyond $N= 600$, the computation time, $t$, follows the power law $t \propto N ^{2.87}.$ This reflects the fact that the computation time becomes dominated by the nested matrix multiplications. Naively matrix multiplications scale as ${\mathcal O} \left( N^3 \right)$ because all $N^2$ elements of the first matrix must be multiplied by $N$ elements in the second matrix. Our code does slightly better and scales as ${\mathcal O} \left( N^{2.87} \right)$ because it uses the highly optimized {\tt numpy.dot} routine.
\par It was shown in \cite{taylor2018preparing} that a resolution of $N= 400$ is sufficient to capture nearly all the lensing kernel and power spectrum information. Meanwhile a resolution of $N = 2000$ is required to capture $80 \%$ of the information when using the spherical Bessel weight and an extremely high resolution of $N = 5000$ is needed to capture $97 \%$ of the information for this choice of weight \cite{taylor2018preparing}.

\subsection{Impact of the Flat Universe Approximation on Lensing Spectra}

\begin{figure}
\centering
\includegraphics[width=85mm]{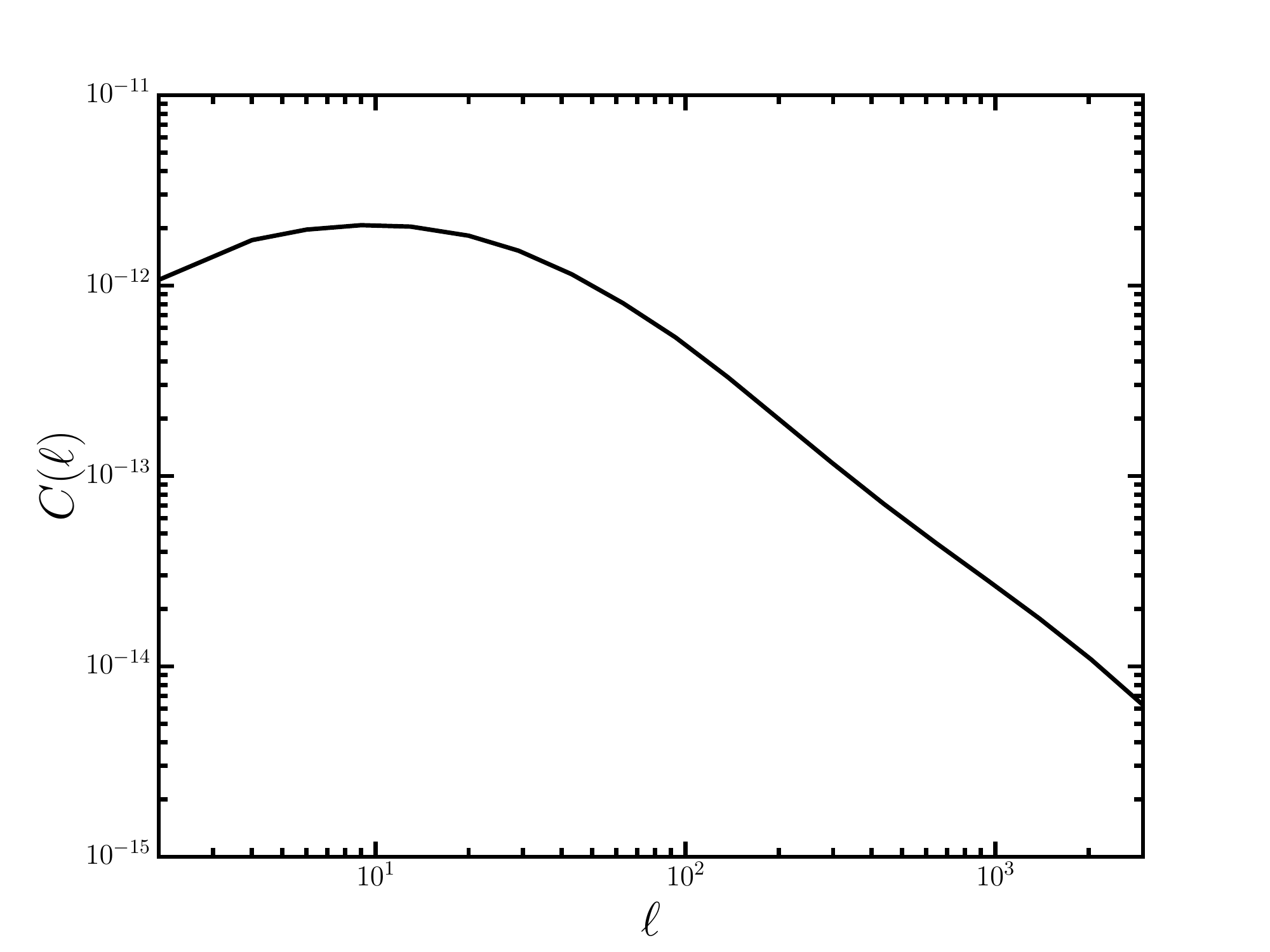}
\includegraphics[width=85mm]{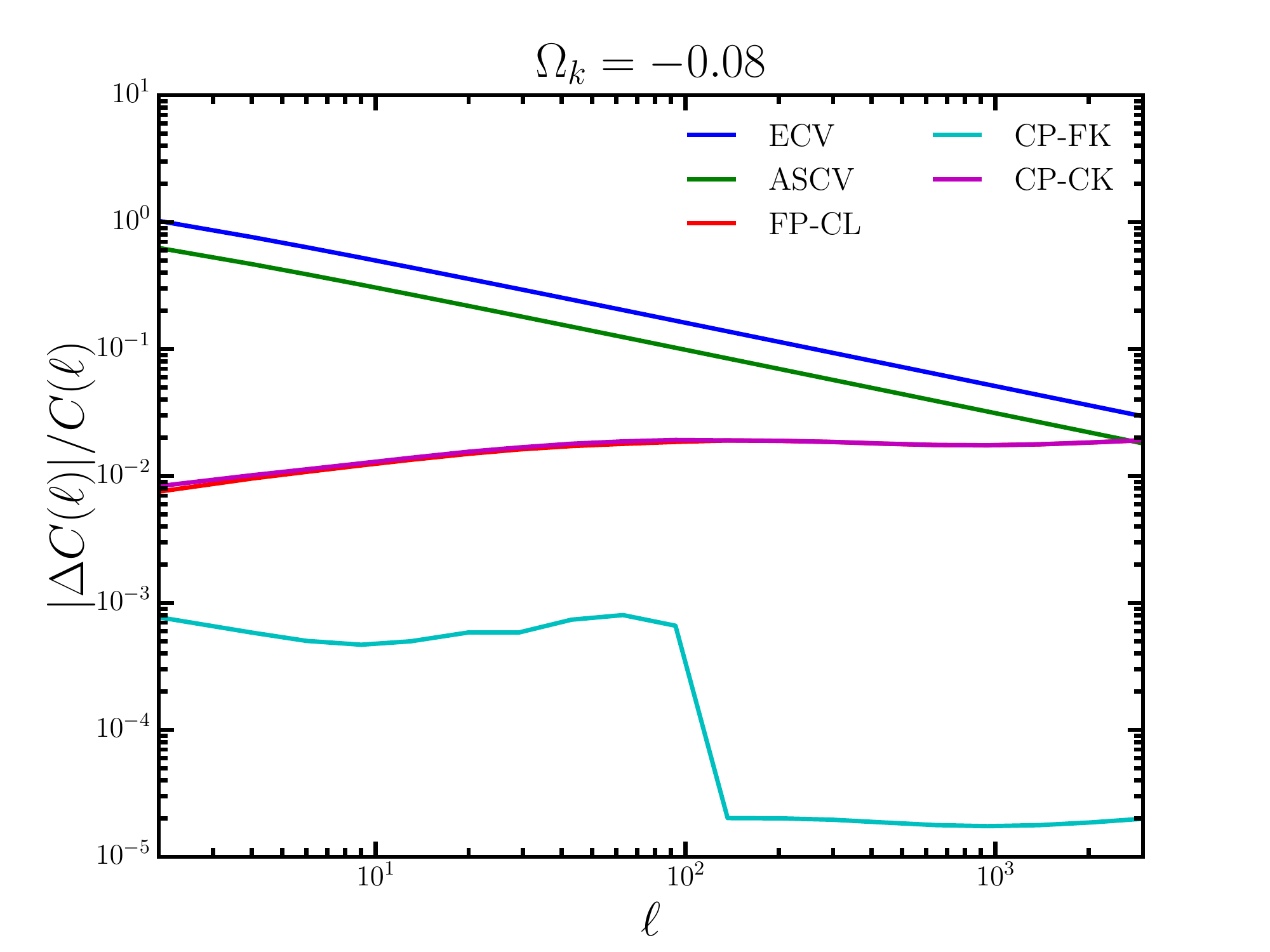}
\includegraphics[width=85mm]{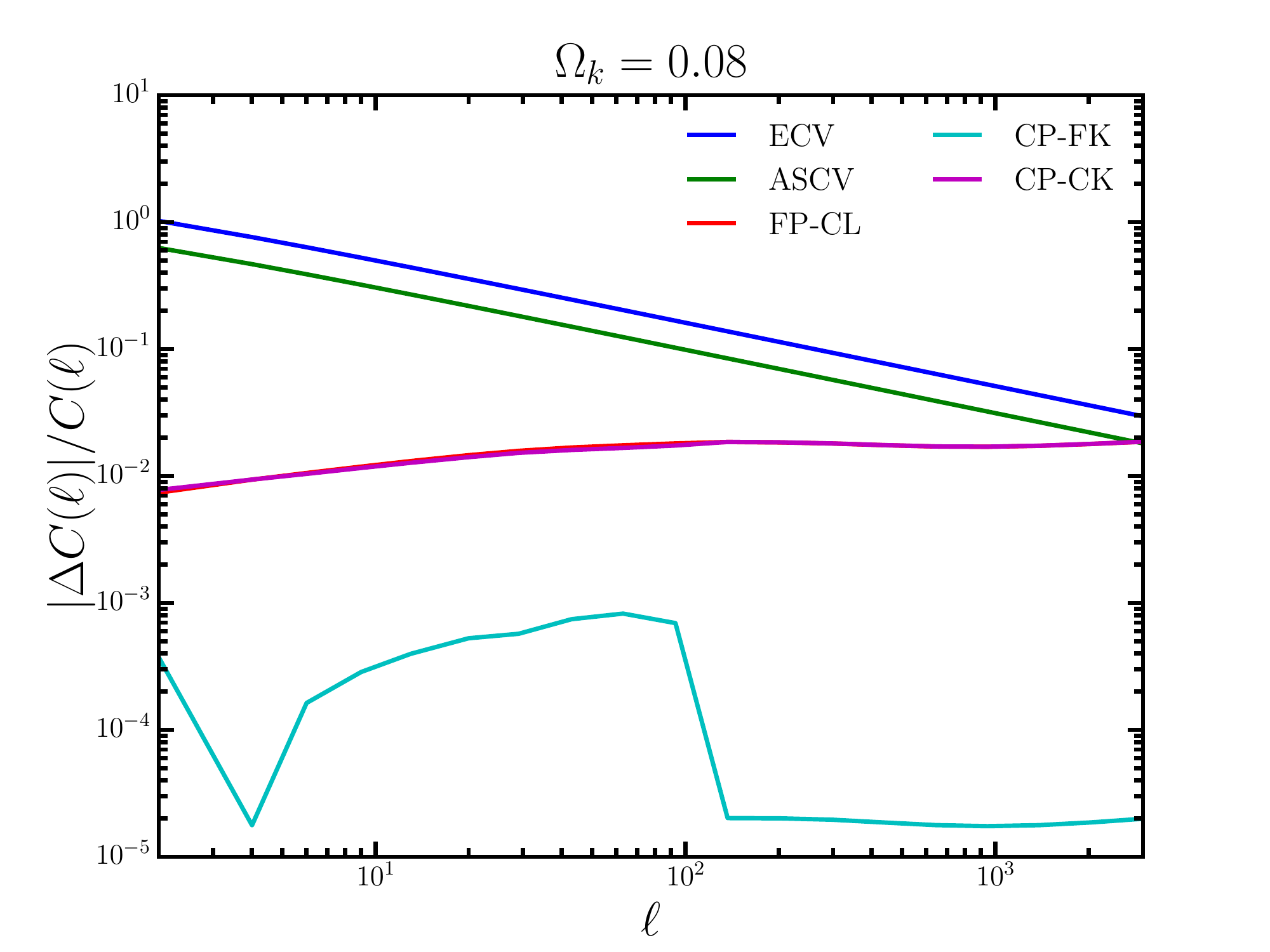}
    \caption{{\bf Top}: The cross-correlated lensing spectrum, $C_\ell$, between tomographic bins $9$ and $10$, using the spatially-flat kernels. {\bf Middle}: Relative change in $C_\ell$ for different kernels with $\Omega_k = -0.08$. The fiducial spectrum, $C_\ell$, uses the spatially-flat kernels. CP, FP, CL and FL respectively denote when the spatially-curved projection kernel, spatially-flat projection kernel, spatially-curved lensing kernel and spatially-curved lensing kernel, are used. The sample variance for Euclid-like (ECV) and the cosmic variance for a theoretical all sky survey (ASCV) are also shown. The relative change in $C_\ell$ is smaller than the sample variance for a Euclid-like survey for all $\ell < 3000$.  {\bf Bottom}: Same as the middle figure, but with $\Omega_k = 0.08$.}
    \label{fig:delta_c_l}
    \end{figure}

\par We compute the cosmic shear power spectra for a flat fiducial cosmology with: $\left( \Omega_m, \Omega_k, \Omega_b,  h_0, n_s, A_s,  \tau  \right) = \left(0.315,\text{ }  0.0,\text{ } 0.04,\text{ }  0.67,\text{ } 0.96,\text{ } 2.1 \times 10 ^9,\text{ } 0.08 \right)$. The linear power spectrum is generated using {\tt CAMB} \cite{camb} and the non-linear part is generated using {\tt HALOFIT} \cite{Halofit}. The Limber approximation is assumed for $\ell > 100.$ The resulting cross-correlated lensing power spectrum between the highest redshift bins is shown at the top of Figure~\ref{fig:delta_c_l}. 
\par For $|\Omega_k| = 0.08$, which is the expected $1 \sigma$ constrain for a Euclid-like experiment~\cite{heavens2006measuring}, we have computed the lensing spectra inside the same bin using spatially-flat lensing and projection kernels (FP-FL). The same lensing spectra using: a spatially-curved projection kernel and spatially -flat lensing kernel (CP-FL), spatially-flat projection kernel and spatially-curved lensing kernel (FP-CL) and  spatially-curved projection and lensing kernels (CP-CL) are computed. The relative difference, $ \Delta C \left( \ell \right) / C \left( \ell \right)$,  between the FP-FL spectrum and the others is shown in the bottom two panels of Figure~\ref{fig:delta_c_l}.  When the spatially-curved projection kernel is used, we employ the modified Limber approximation defined in~\eqref{eq:modlimber} for $\ell > 100$. \footnote{Since the modified Limber approximation in equation~\ref{eq:modlimber} improves at larger $\ell$ \cite{lesgourgues2014fast}, if the relative change in $C_{100}$ due to this approximation is smaller than the samples variance at the largest $\ell$-mode considered, it is negligible across all $\ell$-modes. We have verified that this is the case.}
\par The sample variance for a Euclid-like survey is also shown in Figure~\ref{fig:delta_c_l} and is given by:
\begin{equation}
\Delta C \left( \ell \right) / C \left( \ell \right) = \sqrt{2} \left[ f_ {sky} \left( 2 \ell + 1 \right)\right] ^ {-1/2},
\end{equation}
where $f_ {sky}$ is the fraction of the sky observed by the survey~\cite{weinberg2008cosmology}.
\par In all cases we find that the relative difference between the spectra computed using the spatially-flat projection and lensing kernels are smaller than the sample variance, up to $\ell = 3000$. This is true for the cross-correlation between all tomographic bins. Since the majority of the information from upcoming cosmic shear studies will be extracted from $\ell$-modes below $\ell = 3000$\cite{taylor2018preparing}, and the impact of changing the projection and lensing kernel is sub-dominant to the sample variance at these scales, it is safe to make the Spatially-Flat Universe Approximation for stage IV experiments. 
\par Testing the impact of the Spatially-Flat Universe approximation with $|\Omega_k| = 0.08$ was an extremely conservative choice.  The 2015 Planck $2 \sigma$ multi-probe constrain place $\left| \Omega_k \right| < 0.005$ \cite{ade2016planck}. We have computed the impact of the Spatially-flat Universe Approximation in this case with  $\left| \Omega_k \right| = 0.005$ and found that relative difference, $\Delta C \left( \ell \right) / C \left( \ell \right)$,  between the CP-CL and FP-FL falls a further order of magnitude from the $|\Omega_k| = 0.08$ case shown in Figure~\ref{fig:delta_c_l}.

\section{Conclusion}
  We have presented the {\tt GLaSS} code that computes generalized cosmic shear power spectra. 
Spherical Bessel and tomographic lensing spectra with an equal number of galaxies per bin and equal redshift run-mode options are available. 
More generally {\tt GLaSS} is capable of computing the lensing spectra with \emph{any} data weighting. This should prove useful for determining the optimal weight for shear data in upcoming surveys.
\par {\tt GLaSS} is fast. Using the Limber approximation, {\tt GLaSS} can compute a $10$-bin tomographic lensing spectra for a single cosmology, sampling $50$ $\ell$-modes, in less than $0.4 s$. For Stage IV experiments where the Limber approximation must be dropped below $\ell <100$, the same spectra is computed in $1.3 s$.
\par Using {\tt GLaSS} we have tested the Spatially-Flat Universe Approximation, which is implicitly assumed in all cosmic shear studies to date.
We find this is an accurate approximation and it is unnecessary to compute the full expression for upcoming surveys.

\section*{Acknowledgements}
We thank the {\tt Cosmosis} team for making their code publicly available and the anonymous referee whose comments significantly improved the paper. PT is supported by the UK Science and Technology Facilities Council. TK is supported by a Royal Society University Research Fellowship.
The authors acknowledge the support of the Leverhume Trust.

\bibliographystyle{apsrev4-1.bst}
\bibliography{bibtex.bib}

\end{document}